\documentclass[3p,times,number]{elsarticle}

\usepackage{mathtools}
\usepackage{lineno,hyperref}
\usepackage{cleveref}
\usepackage{color,xspace}
\usepackage{adjustbox}
\usepackage{booktabs}
\usepackage{comment}
\usepackage{footnote}
\usepackage{graphicx}
\makesavenoteenv{tabular}

\modulolinenumbers[5]

\newcommand{\muec}{$\mu$--$e$ conversion\xspace}
\newcommand{\neqflux}{\mathrm{n_{eq}\,cm^{-2}}}

\begin{document}

\begin{frontmatter}

\title{Radiation hardness study for the COMET Phase-I electronics}

\author[osaka]{Yu~Nakazawa\corref{cor}}
\ead{y-nakazawa@kuno-g.phys.sci.osaka-u.ac.jp}
\author[monash]{Yuki~Fujii}
\author[icl]{Ewen~Gillies}
\author[kek]{Eitaro~Hamada}
\author[kek]{Youichi~Igarashi}
\author[kaist]{MyeongJae~Lee}
\author[kek]{Manabu~Moritsu}
\author[osaka]{Yugo~Matsuda}
\author[kyushu]{Yuta~Miyazaki}
\author[kyushu]{Yuki~Nakai}
\author[kaist]{Hiroaki~Natori\fnref{hiro}}
\author[kyushu]{Kou~Oishi\fnref{kou}}
\author[osaka]{Akira~Sato}
\author[icl]{Yoshi~Uchida}
\author[kek]{Kazuki~Ueno}
\author[kek]{Hiroshi~Yamaguchi}
\author[kaist]{BeomKi~Yeo}
\author[osaka]{Hisataka~Yoshida}
\author[ihep]{Jie~Zhang}

\cortext[cor]{Corresponding author}
\fntext[hiro]{Present address: Institute of Materials Structure Science, High Energy Accelerator Research Organization, Tsukuba, Ibaraki 305-0801, Japan}
\fntext[kou]{Present address: Department of Physics, Imperial College London, London SW7 2AZ, UK}

\address[osaka]{Department of Physics, Graduate School of Science, Osaka University, 1-1 Machikaneyama-cho, Toyonaka, Osaka 560-0043, Japan}
\address[monash]{School of Physics and Astronomy, Monash University, Clayton, Victoria 3800, Australia}
\address[icl]{Department of Physics, Imperial College London, London SW7 2AZ, UK}
\address[kek]{Institute of Particle and Nuclear Studies, High Energy Accelerator Research Organization, 1-1 Oho, Tsukuba, Ibaraki 305-0801, Japan}
\address[kaist]{Center for Axion and Precision Physics Research, Institute for Basic Science, Daejeon 34051, South Korea}
\address[kyushu]{Department of Physics, Kyushu University, 744 Moto-oka, Nishi-ku, Fukuoka 819-0395, Japan}
\address[ihep]{State Key Laboratory of Particle Detection and Electronics, Institute of High Energy Physics, CAS, Beijing 100049, China}

\begin{abstract}
Radiation damage on front-end readout and trigger electronics is an important issue in the COMET Phase-I experiment at J-PARC, which plans to search for the neutrinoless transition of a muon to an electron.
To produce an intense muon beam, a high-power proton beam impinges on a graphite target, resulting in a high-radiation environment.
We require radiation tolerance to a total dose of 1.0\,kGy and 1\,MeV equivalent neutron fluence of $1.0\times10^{12}\,\mathrm{n_{eq}\,cm^{-2}}$ including a safety factor of 5 over the duration of the physics measurement.
The use of commercially-available electronics components which have high radiation tolerance, if such components can be secured, is desirable in such an environment.
The radiation hardness of commercial electronic components has been evaluated in gamma-ray and neutron irradiation tests.
As results of these tests, voltage regulators, ADCs, DACs, and several other components were found to have enough tolerance to both gamma-ray and neutron irradiation at the level we require.
\end{abstract}


\end{frontmatter}

\section{Introduction}
The COherent Muon To Electron Transition (COMET) Phase-I experiment will take place at the Japan Proton Accelerator Research Complex (J-PARC) in Tokai, Japan~\cite{COMET:Proceedings}.
The aim of this experiment is to search for the neutrinoless transition of a muon to an electron (\muec) in a muonic atom, which has never been observed yet~\cite{mu-e}, with a single event sensitivity of $3\times10^{-15}$ in a 150 day-long physics measurement.
This sensitivity is a factor of $100$ better than the current limit from SINDRUM-II~\cite{SINDRUM-II}.
In order to reach this sensitivity, an $8\,\mathrm{GeV}$, $3.2\,\mathrm{kW}$ proton beam is extracted from the J-PARC Main Ring and impinged on a graphite target; this will enable to produce the most intense muon beam produced from pion decay.
The muons are stopped in aluminium target disks and form muonic atoms.
The signal electrons from the \muec are detected by a cylindrical drift chamber and a set of trigger hodoscope counters in a solenoid magnet.
To measure the muon beam profile and the beam background, we also adopt straw-tube trackers and electron calorimeters, which are prototype detectors for the COMET Phase-II experiment.
We have developed several types of readout and trigger electronics for recording the data from these detectors.

\renewcommand{\thefootnote}{\fnsymbol{footnote}}
Interaction of a high-power proton beam with the pion-production target results in a high-radiation environment in the experimental hall.
Since the front-end and signal processing electronics are located near the detector to avoid effects such as electrical noise, cross-talk, signal degradation, they are greatly affected by the radiation environment.
In particular, the total ionizing dose (TID) from gamma-rays and the displacement damage dose (DDD) from neutrons cause significant degradation in the performance of electronics.
In order to understand the radiation environment around the COMET beamline, independent simulation studies using PHITS version 2.76~\cite{phits} and Geant4 version 10.1~\cite{geant4} were performed.
According to the simulation study, the electronics are expected to withstand a total dose of $1.0\,\mathrm{kGy}$ and neutron fluence of $1.0\times10^{12}\,\neqflux$ \footnote[4]{$\neqflux$ is the unit of $1\,\mathrm{MeV}$ equivalent neutron fluence.} during the 150-day physics data taking period~\cite{COMET:Radiation}, where radiation levels include a safety factor of $5$ to take into account the uncertainties in the simulation results.

To ensure that the electronics work stably at these radiation levels, we have investigated the radiation hardness of the commercial electronic components that would be used in the COMET front-end electronics and most importantly the power regulator, analog-to-digital converter (ADC) and digital-to-analog converter (DAC).
The development of a new radiation-hard application specific integrated circuit (ASIC) is not an option on the time-scale of the COMET Phase-I experiment.
Commercial high-reliability electronic components for space and military application also exist for some parts, however, these are overly-specialised for the purposes of this experiment.
Such high-reliability parts do not exist in the case of electronic components characterized by high speed, high efficiency, and low noise, which are all necessary specifications in the development of the COMET front-end electronics boards.
Therefore, the choice was made to use more standard components, but required that each of the components that are employed in the front-end board are evaluated on their radiation hardness using gamma-ray sources and neutron beams.
In this paper, the results of these tests are summarized.\footnote[3]{All authors declare that: (i) no support, financial or otherwise, has been received from any chip company; and (ii) there are no other relationship or activities that could appear to have influenced the submitted work.}

\section{Electronics components for COMET Phase-I}
All the electronic components are required to have enough radiation tolerance in the COMET Phase-I experiment that the readout electronics can be operated stably during the whole physics data-taking run.
Soft and hard errors on the components need to be investigated.
Field-programmable gate arrays (FPGAs) are widely used in the front-end electronics boards, but soft errors by neutrons can generate bit flips.
We have developed FPGA firmware which incorporate auto-recovery schemes to repair these flips.
The error rates for several FPGAs have already been tested and reported~\cite{Neutron:FPGA}.
Compared to the soft errors, hard damage from TID and DDD to electronic components are permanent and unrecoverable.
In this study, commercial parts which are not certified for radiation-hardness were tested.
The electronic components that were tested for radiation hardness include voltage regulators, Digital to Analog Converters (DAC), Analog to Digital Converters (ADC), splitters, OR gates, and multiplexers.
Several different brands and types of regulators are tested, as the required specification on the operating voltages are different from one board to another.
Switching regulators generate lower heat than linear regulators as a result of the high power efficiency and are suitable for use in temperature-controlled areas and for the first stage when regulating external power.
On the other hand, linear regulators allow for simpler circuit configurations and lower noise levels.
We investigated the radiation tolerance for both types of regulator.
Analog circuits must be used to process the analog signals from the detectors.
DACs are used to adjust their processing parameters such as threshold voltages.
To improve the signal-to-noise ratio, all the readout electronics which have on-board ADC chips are located close to the detectors.
In this situation, it is essential to implement ADCs on the readout electronics.
Splitters, OR gates, and multiplexers are needed to distribute a common clock and a trigger signal to all the front-end electronics.
In the COMET trigger system, differential signal lines are used for those signals, and single-end signal lines are also used in laboratory tests.

\section{Experimental setup}
The experimental setup for each electronic components is presented in the following.
Candidates for the electronic components are summarized in Table~\ref{table:EvalIc}.
In the tests, only the target samples were located in the radiation area.
To avoid radiation damage to other devices such as power supplies and data loggers, they were located on the outside of the radiation area.

\begin{table}[htb!]
\centering
\caption{Candidates for the front-end electronics in the COMET Phase-I experiment.}
\vspace{3mm}
\label{table:EvalIc}
\begin{tabular}{cc}
 \begin{minipage}{0.3\textwidth}
  \begin{center}
  \scriptsize
  \begin{tabular}{|p{23mm}@{\hspace{1pt}}p{12mm}|} \hline
  Manufacturer          & Name       \\ \hline
  \multicolumn{2}{|c|}{\shortstack{\bf{Positive-linear regulator} \rule[-2mm]{0mm}{5mm} }}\\ 
  Linear Technology     & LT1963     \\ 
                        & LT1963-3.3 \\
                        & LT1963-2.5 \\
                        & LT1963-1.8 \\
                        & LT3070     \\
                        & LT1764A    \\
                        & LTC3026    \\ 
  Maxim Integrated      & MAX8556    \\
  Texas Instruments     & TPS7A7200  \\
                        & TPS75801   \\
                        & TPS74401   \\
  Analog Devices        & ADP1755    \\ \hline
  \multicolumn{2}{|c|}{\shortstack{\bf{Positive-switching regulator} \rule[-2mm]{0mm}{5mm} }} \\
  Linear Technology     & LT8612     \\ 
                        & LT8614     \\
                        & LTM4620    \\
                        & LTM4644    \\
                        & LTM8033    \\ 
  Texas Instruments     & LMZ10503   \\ \hline
  \multicolumn{2}{|c|}{\shortstack{\bf{Negative-linear regulator} \rule[-2mm]{0mm}{5mm} }}\\ 
  ST Microelectronics   & L79        \\
  ON Semiconductor      & MC7905     \\
  New JRC               & NJM2828    \\
  Linear Technology     & ADP7182    \\ 
                        & LT1964     \\
                        & LT3015     \\
                        & LT3032     \\
                        & LT3090     \\ 
                        & LT3091     \\ 
  Texas Instruments     & LM337      \\
  Microchip Technology  & MIC5271    \\
                        & TC59       \\ \hline
  \end{tabular}
  \end{center}
 \end{minipage}
 \begin{minipage}{0.3\textwidth}
  \begin{center}
  \scriptsize
  \begin{tabular}{|p{19mm}@{\hspace{4pt}}p{16mm}|} \hline
  Manufacturer      & Name          \\ \hline
  \multicolumn{2}{|c|}{\shortstack{\bf{DAC} \rule[-2mm]{0mm}{5mm} }}\\ 
  Linear Technology & LTC2624       \\ 
                    & LTC2634       \\
                    & LTC2654       \\
  Texas Instruments & DAC7564       \\
                    & DAC7565       \\
  Analog Devices    & AD5624R       \\
                    & AD5684R       \\ 
                    & AD5324        \\
                    & AD5624        \\
                    & AD5684        \\ \hline
  \multicolumn{2}{|c|}{\shortstack{\bf{ADC} \rule[-2mm]{0mm}{5mm} }}\\
  Linear Technology & LTC2264       \\ 
  Analog Devices    & AD9287        \\
                    & AD9637        \\ \hline
  \multicolumn{2}{|c|}{\shortstack{\bf{High-Speed Differential Line Driver} \rule[-2mm]{0mm}{5mm} }}\\
  Texas Instruments & SN65LVDS391   \\
                    & SN65LVDS116   \\
                    & SN65LVDS386   \\ \hline
  \multicolumn{2}{|c|}{\shortstack{\bf{High-Speed Differential Receiver} \rule[-2mm]{0mm}{5mm} }}\\
  Texas Instruments & SN65LVDT348   \\ \hline
  \multicolumn{2}{|c|}{\shortstack{\bf{Splitter} \rule[-2mm]{0mm}{5mm} }}\\
  Maxim Integrated  & MAX9175       \\ \hline
  \multicolumn{2}{|c|}{\shortstack{\bf{Positive OR gate} \rule[-2mm]{0mm}{5mm} }}\\
  Texas Instruments & SN74AUP1G32   \\ \hline
  \multicolumn{2}{|c|}{\shortstack{\bf{NOR/OR gate} \rule[-2mm]{0mm}{5mm} }}\\
  Texas Instruments & CD4078BM96    \\ \hline
  \multicolumn{2}{|c|}{\shortstack{\bf{Multiplexer} \rule[-2mm]{0mm}{5mm} }} \\
  Analog Devices    & ADG1606       \\ \hline
  \end{tabular}
  \end{center}
  \normalsize
 \end{minipage}
 \end{tabular}
 \vspace*{10pt}
\end{table}


%
\begin{figure}
 \vspace{-5mm}
 \centering
 \begin{minipage}{0.29\textwidth}
  \centering
  \includegraphics[width=\textwidth]{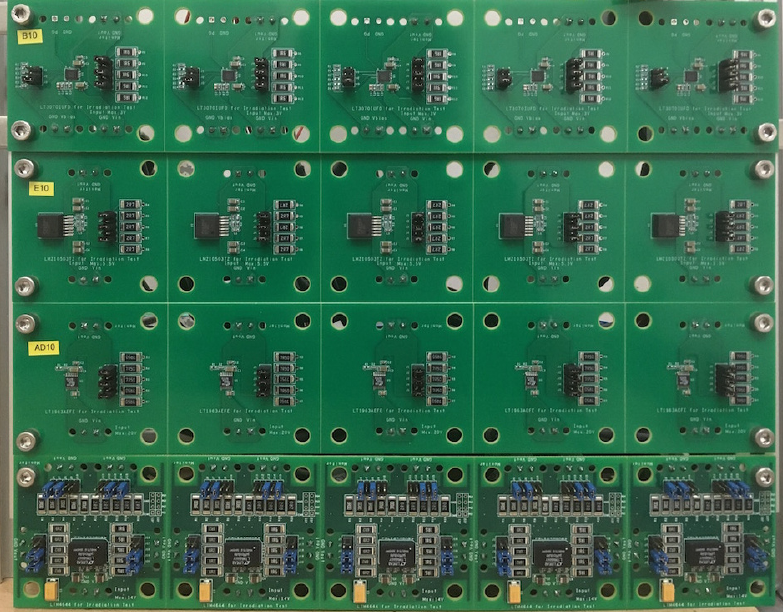}\\
  (a)
  \label{fig:CustomBoards}
 \end{minipage}
 \begin{minipage}{0.18\textwidth}
  \vspace{9mm}
  \centering
  \includegraphics[width=\textwidth]{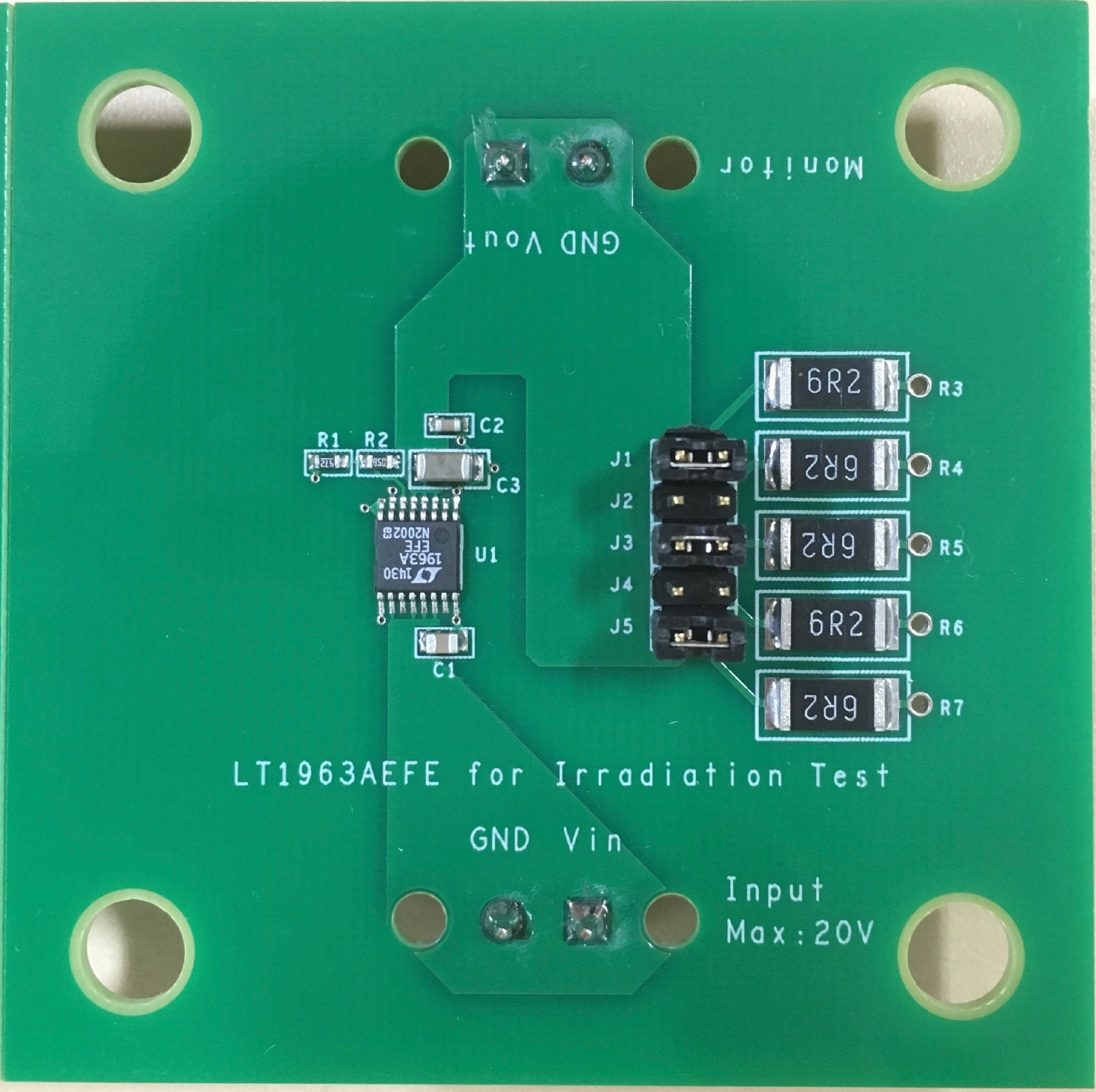}\\
  (b)
  \label{fig:ACustomBoard}
 \end{minipage}
 \vspace{-1mm}
 \caption{The custom PCB boards for positive-voltage regulator test. (a) One set of custom test boards. There were more than five samples for each positive-voltage regulator. (b) A custom board for LT3070. Resistors were also mounted on the test board.}
 \label{fig:SEUrate}
\end{figure}
%

\subsection{Voltage regulators}
To investigate the radiation hardness of commercial voltage regulators, we produced custom printed circuit boards (PCB) as shown in Figure~\ref{fig:SEUrate}.
For the other regulators, we used commercial evaluation boards provided by the regulator manufacturers.
The conditions of the real front-end readout and trigger boards are emulated by attaching load resistors to the regulators.
The prototypes of the readout electronics developed for the COMET Phase-I experiment give a realistic current load environment and were utilized to test a subset of the regulators.
The following commercial evaluation boards were used;
\begin{itemize}
  \item Linear Technology 
    \begin{description}
      \item[] DC2010A for LT8612
      \item[] DC2019B for LT8614
      \item[] DC1623A for LTM8033
    \end{description}
  \item Texas Instruments 
    \begin{description}
      \item[] TPS7A7200EVM-718 for TPS7A7200
      \item[] TPS74401EVM-118 for TPS74401
    \end{description}
\end{itemize}

\begin{figure}[htb!]
 \begin{center}
 \includegraphics[width=0.6\textwidth]{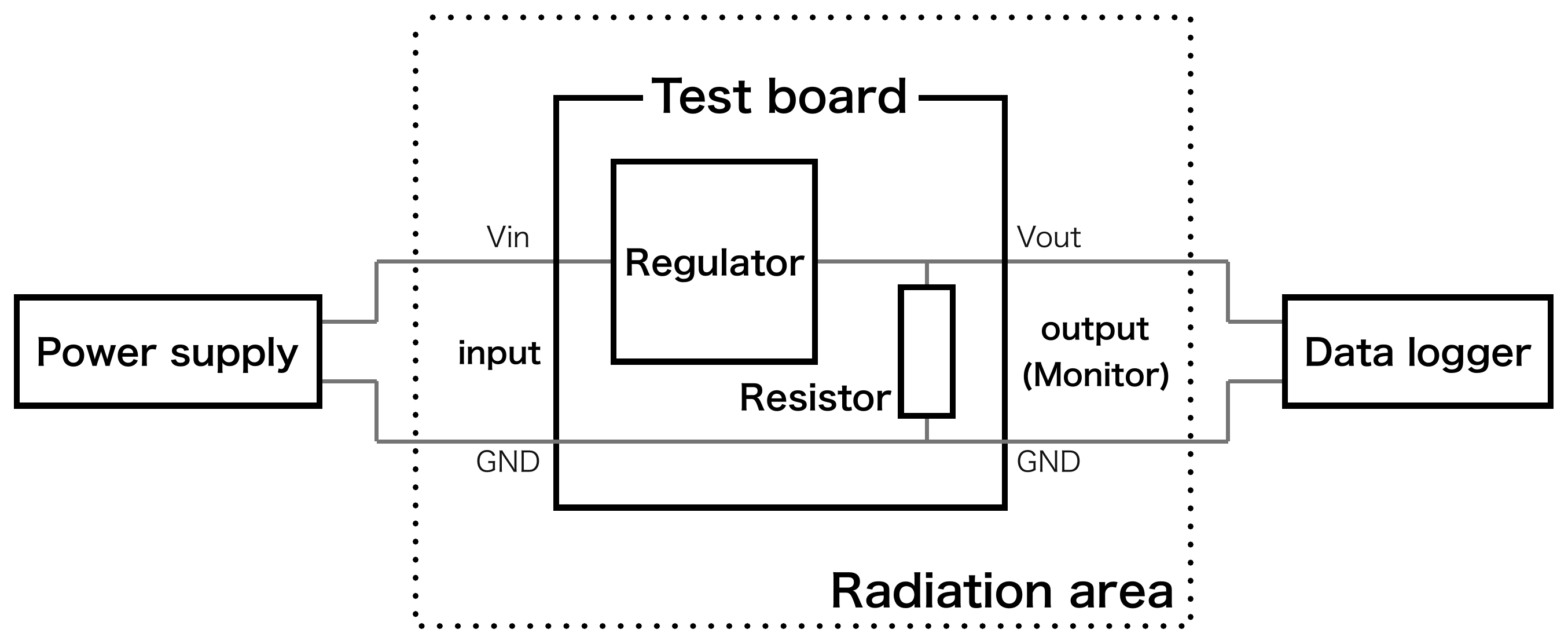}
 \end{center}
 \vspace{-5mm}
 \caption{The schematic view of the experimental setup. Voltage regulators were supplied with input voltage and their outputs were recorded by data loggers outside a radiation area.}
 \label{fig:Setup}
\end{figure}

Figure~\ref{fig:Setup} shows the typical experimental setup.
Only the test boards are located in the radiation area.
Power supply modules and data loggers were located outside the radiation area to avoid radiation damage.
The output voltage of the regulators was monitored and recorded by data loggers during exposure.
%
%
The gamma-ray irradiation dose rate was chosen to be $4.5\,\mathrm{Gy\,h^{-1}}$ which is 80 times higher than the rate of the COMET Phase-I experiment, $6.0\times10^{-2}\,\mathrm{Gy\,h^{-1}}$.
Tests under higher dose rates were also performed for some regulators in order to investigate the dependence of the gamma-ray tolerance on the dose rate.
These dose rate were $2.2\times10^1\,\mathrm{Gy\,h^{-1}}$, $2.0\times10^2\,\mathrm{Gy\,h^{-1}}$, $2.3\times10^2\,\mathrm{Gy\,h^{-1}}$, and $4.0\times10^2\,\mathrm{Gy\,h^{-1}}$.
The 29 types of voltage regulators listed in Table~\ref{table:EvalIc} were irradiated with gamma-rays.
To obtain reliable results for the electronics used in the COMET Phase-I experiment, we irradiated six or more samples for each type of regulator except LT8612 and LT8614, for which only three samples each were tested.
Power cycles were performed every $2.0\times10^2\,\mathrm{Gy}$ exposure because it was observed that voltage regulators could break when performing a power cycle after irradiation.

\subsection{DACs}
For the selection of radiation-hard DACs, ten types of DAC were tested.
DAC candidates were mounted on handmade circuit boards for both neutron and gamma-ray irradiations.
Before irradiation, the register values on the DACs were determined.
A fixed input voltage was applied to a DAC during irradiation.
The DAC output signals were measured and recorded with a data logger.

\subsection{ADCs}
Several test PCB boards with ADCs were fabricated and used for radiation tests.
In the gamma-ray irradiation tests, the ADCs received periodical square waves from a function generator and generated digitized waveforms during irradiation.
Differences in the waveforms recorded before and during irradiation were used to evaluate the deterioration from radiation.
In the case of neutron irradiation, the prototype boards of the readout electronics on which the ADCs are to be implemented were irradiated.

\subsection{Other components}
\begin{figure}[htb!]
 \begin{center}
 \includegraphics[width=0.3\textwidth]{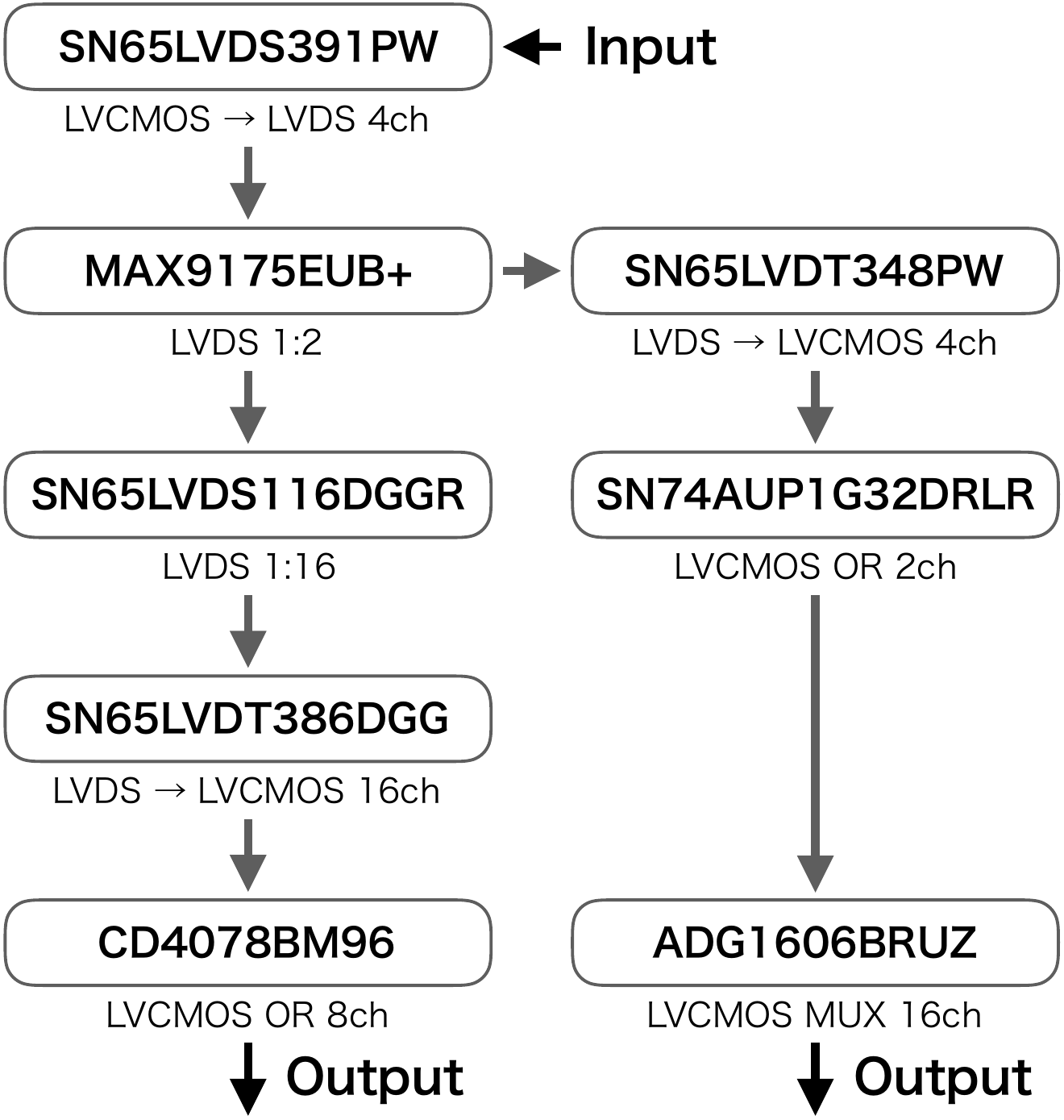}
 \end{center}
 \vspace{-5mm}
 \caption{Block diagram of the test board for the components selection.}
 \label{fig:ROESTI_IF}
\end{figure}

We performed an integrated irradiation test on the electronics components listed below.
The circuit diagram of the test board developed in this purpose is shown in Figure~\ref{fig:ROESTI_IF}.
\begin{itemize}
  \item High-Speed Differential Line Driver: SN65LVDS391, SN65LVDS116 and SN65LVDS386
  \item High-Speed Differential Receiver: SN65LVDT348
  \item Splitter: MAX9175
  \item Positive OR gate: SN74AUP1G32
  \item NOR/OR gate: CD4078BM96
  \item Multiplexer: ADG1606
\end{itemize}
The test board has an input port for an SN65LVDS391 line driver, which converts a single-end signal to differential signals and sends them to MAX9175.
Outputs of the MAX9175 are sent to SN65LVDT348 converting them to single-end signals and SN65LVDS116 dividing them to 16 differential lines.
Two lines of the four single-end signals from SN65LVDT348 are passed through the OR gate of SN74AUP1G32 and the multiplexer of ADG1606.
On the other hand, all the differential signals from SN65LVDS116 are converted to single-end signals by SN65LVDT386 and processed by the logical disjunction in CD4078BM96.
During irradiation, a pulsed input signal was injected into SN65LVDS391 and the outputs from ADG1606 and CD4078BM96 were measured by using an oscilloscope.
This evaluation method was adopted for both gamma-ray and neutron irradiation tests.

\section{Experimental facilities}
The gamma-ray irradiation tests were performed at four facilities: the Cobalt-60 Gamma-ray Irradiation Facility at the National Institute for Quantum and Radiological Science and Technology, formerly known as JAEA Takasaki; the Tokyo Institute of Technology Radioisotope Research Center; Research Laboratory for Quantum Beam Science, Institute of Scientific and Industrial Research (ISIR) at Osaka University; and the Advanced Radiation Technology Institute of the Korea Atomic Energy Research Institute.
All the facilities provide $1.17$ and $1.33\,\mathrm{MeV}$ gamma-rays from $^{60}\rm{Co}$ sources.
Dose rates were controlled by changing the distance between the $^{60}\rm{Co}$ source and the target sample.

For the neutron irradiation, We used two tandem electrostatic accelerators at Kobe University and Kyushu University.
The accelerator in Kobe University generates a neutron beam with an energy peak of $2\,\mathrm{MeV}$ from the $^9\rm{Be\left(d,n\right)}^{10}\rm{B}$ reaction and an incoming $3\,\mathrm{MeV}$ deuteron beam.
The neutron beam flux is $\left(4.9 \pm 1.5\right) \times 10^{6} \,\neqflux \mathrm{\,sec^{-1}}$ with a $1\,\mathrm{\mu A}$ beam current at $10\,\mathrm{cm}$ from the Be target on the beam axis, including distance and angular uncertainties.
Another tandem electrostatic accelerator in Kyushu University generates a neutron beam from the $^{12}\rm{C\left(d,n\right)}^{13}\rm{N}$ reaction with a $9\,\mathrm{MeV}$ deuteron beam.
According to past measurement results in this facility~\cite{Neutron:Kyushu}, the neutron beam flux is $4.3 \times 10^{7} \,\neqflux \mathrm{\,sec^{-1}}$ with a $1\,\mathrm{\mu A}$ beam current at $10\,\mathrm{cm}$ from the C target on the beam axis.

\section{Measurements \& Results}
To investigate the TID effect of gamma-rays and the DDD effect of neutrons on each electronic component with the total dose of 1.0\,kGy and the neutron fluence of $1.0\times10^{12}\neqflux$ respectively, we measured their gamma-ray and neutron tolerances.
\subsection{Voltage regulators} 
\subsubsection{Gamma-ray tolerance}
\begin{figure}[htb!]
 \begin{center}
 \includegraphics[width=0.48\textwidth]{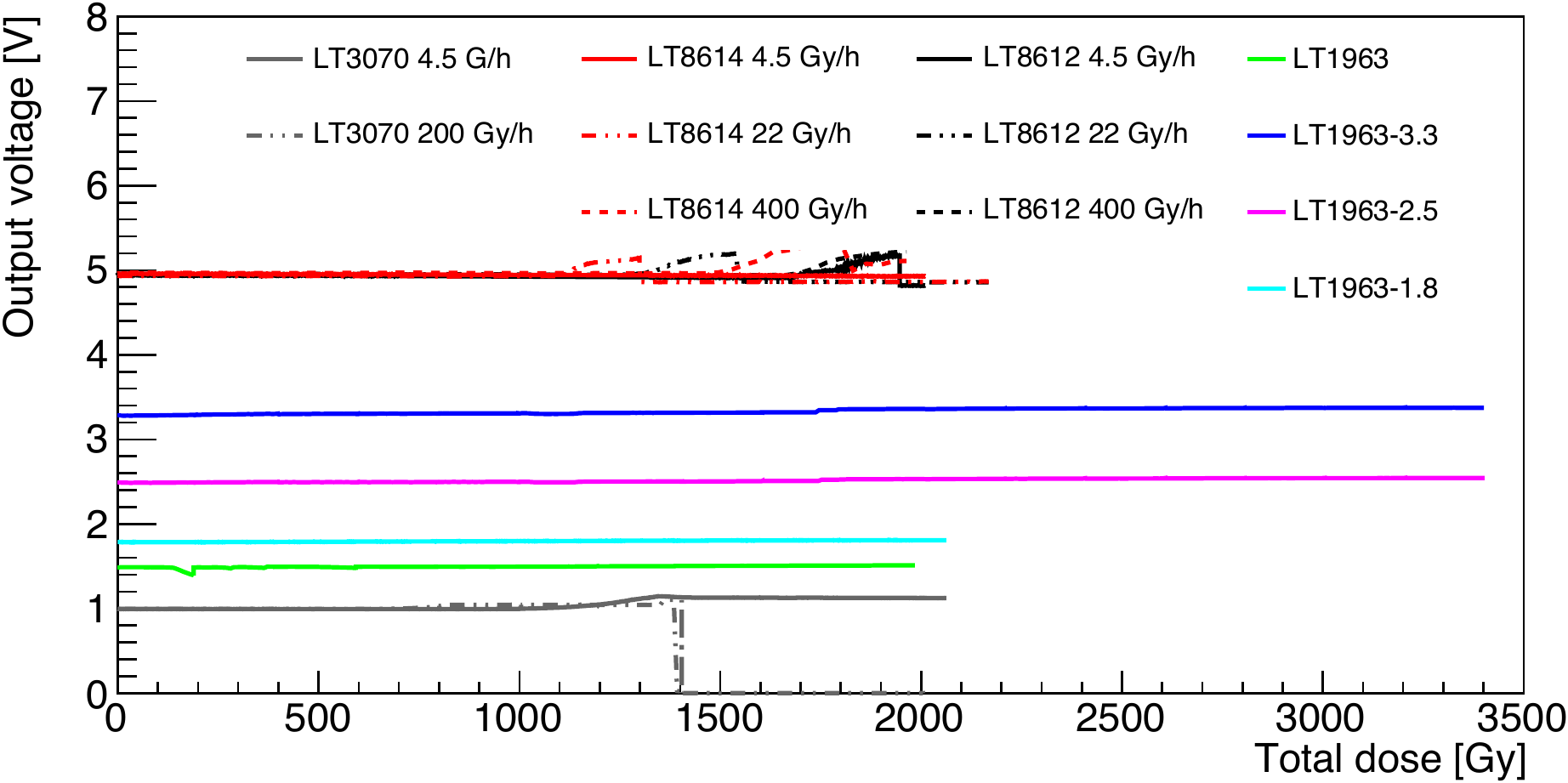}
 \end{center}
 \vspace{-5mm}
 \caption{Results of positive-switching/linear regulators in the gamma-ray irradiation tests.}
 \label{fig:PosiReg}
\end{figure}

Figure~\ref{fig:PosiReg} shows the test results of two of the positive-switching regulators (LT8612 and LT8614) and five of the positive-linear regulators (four LT1963 series and LT3070).
Both positive-switching regulators survived beyond the required dosage.
We evaluated the dose-rate dependence of gamma-ray tolerance using three samples of LT8612 and LT8614.
All those samples show the voltage increasing from 5.0\,V up to 5.4\,V after irradiation of 1.0\,kGy, independently of the dose rate.
These changes of the output were not related with the timing of power cycle either.
Since the increase in their outputs occurred after 1.0\,kGy irradiation, they meet the requirement.
LTM4620, LTM4644, and LMZ10503 were suddenly broke at 0.3\,kGy, 1.1\,kGy and 0.8\,kGy irradiation, respectively.
The output of LTM8033 increased from 3.3\,V to 3.6\,V after 0.2\,kGy irradiation and becomes more than 4.0\,V at 1.0\,kGy irradiation.
%

Among the irradiated positive-linear regulators, all the series of LT1963 survived beyond the required dosage.
In the tests, we set the output of LT1963, whose output is adjustable, to 1.5\,V.
After 0.1\,kGy irradiation, the outputs of all LT1963s dropped.
However, this deterioration was recovered by raising the input voltage and was not observed after 0.6\,kGy irradiation.
Following gamma-ray irradiation, the output of LT1963-3.3 changed from 3.3\,V to 3.4\,V and the output of LT1963-2.5 changed from 2.5\,V to 2.6\,V.
These changes are small enough for the absolute maximum ratings of the other electronic components, to which these regulators apply voltage.
Five samples of LT3070 were exposed with the dose rate of $2.0\times10^2\,\mathrm{Gy\,h^{-1}}$.
Then, all the samples were subject to an increase in output after about 0.7\,kGy irradiation and were broken before 1.5\,kGy irradiation.
After this test, the other five samples of LT3070 were irradiated again with the dose rate of $4.5\,\mathrm{Gy\,h^{-1}}$ and survived beyond 2.0\,kGy irradiation.
The outputs also increased from 1.0\,V to about 1.2\,V, but these phenomena happened after $1.0\,\mathrm{kGy}$ irradiation.
MAX8556, TPS7A7200, TPS74401, TPS70851, and ADP1755 were suddenly broken at 0.8\,kGy, 1.0\,kGy, 1.4\,kGy, 0.6\,kGy, and 0.8\,kGy, respectively.

It was confirmed that the gamma-ray tolerances of LT8612, LT8614, the LT1963 series, and LT3070 satisfied the requirements for the COMET Phase-I experiment.
These results also show that the voltage drop in the voltage regulators decreased from gamma-ray irradiation before complete failure.
The result for LT8612 is consistent with the study by the ATLAS New Small Wheel group~\cite{atlas-new-small-wheel}, but the one for ADP1755 is different from that study.
The ATLAS group used a 220\,MeV proton beam, and its energy deposition on the regulator is different from the case of gamma-ray irradiation.
Furthermore, we do not know the structure and material composition of the regulators that would allow us to calculate the energy loss.
This could be the cause of the discrepancy between their results and ours.

\begin{figure}[htb!]
 \begin{center}
 \includegraphics[width=0.48\textwidth]{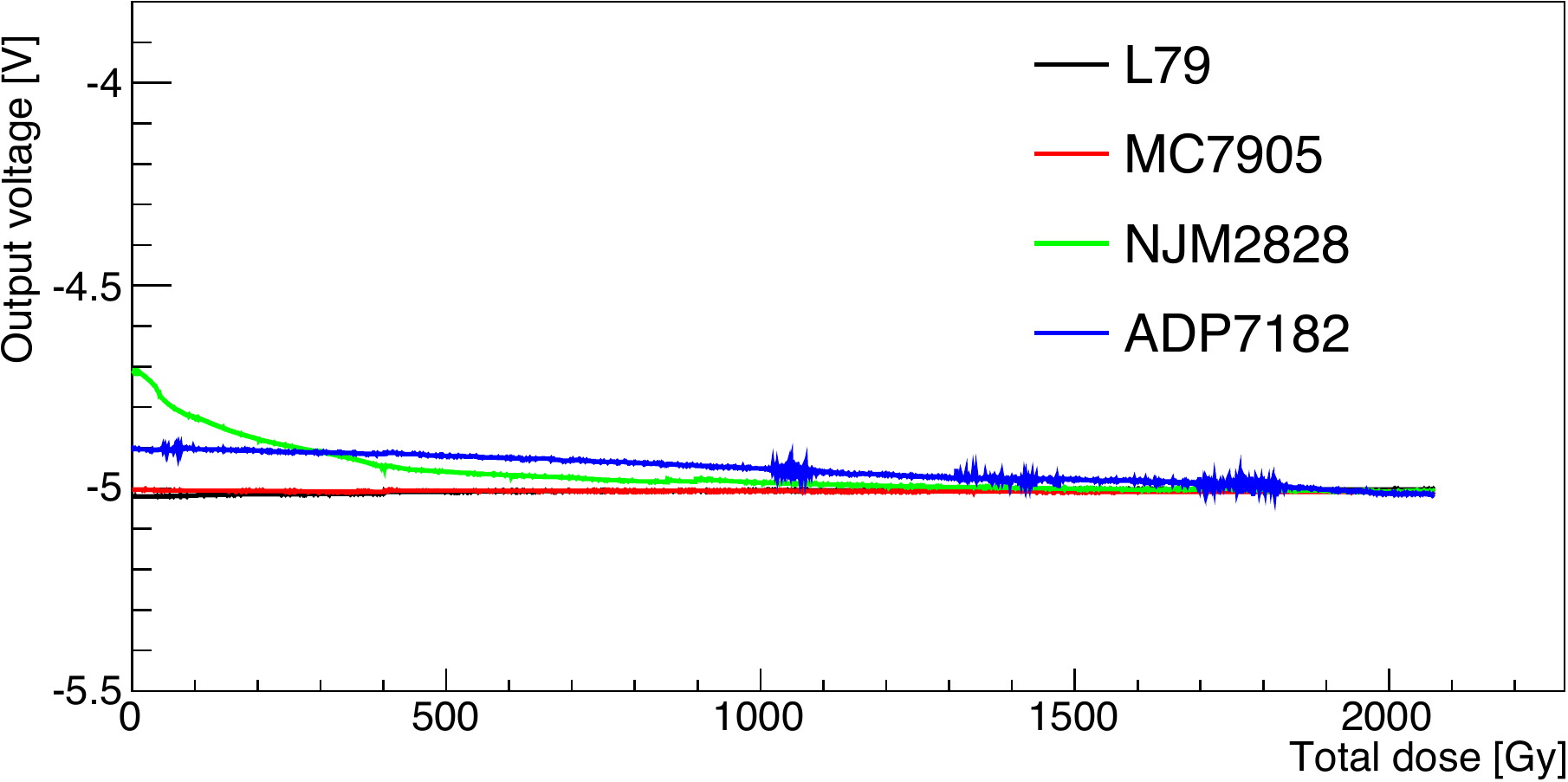}
 \end{center}
 \vspace{-5mm}
 \caption{Result of the gamma-ray irradiation tests for negative-linear regulators.}
 \label{fig:NegaLinear}
\end{figure}

Four negative-linear regulators -- L79, MC7905, NJM2828, and ADP7182 -- survived after 2.0\,kGy irradiation.
Two samples were tested for each regulator.
The outputs of NJM2828 and ADP7182 samples were changed by the TID effect; on the other hand, the outputs of L79 and MC7905 were stable.
A further eight samples for each regulator -- L79 and MC7905 -- were investigated and also survived to 2.0\,kGy irradiation.
The irradiation results of L79, MC7905, and ADP7182 are shown in Figure~\ref{fig:NegaLinear}.
The outputs of LT1964, LT3015, LT3032, LT3090, LT3091, and TC59 were unstable immediately after the irradiation started.
LM337 was suddenly broken at 0.4\,kGy irradiation, and the output of MIC5271 increased from $-5$\,V to 0\,V before 0.2\,kGy irradiation.
%
%

\subsubsection{Neutron tolerance}
\begin{figure}[htb!]
 \begin{center}
 \includegraphics[width=0.48\textwidth]{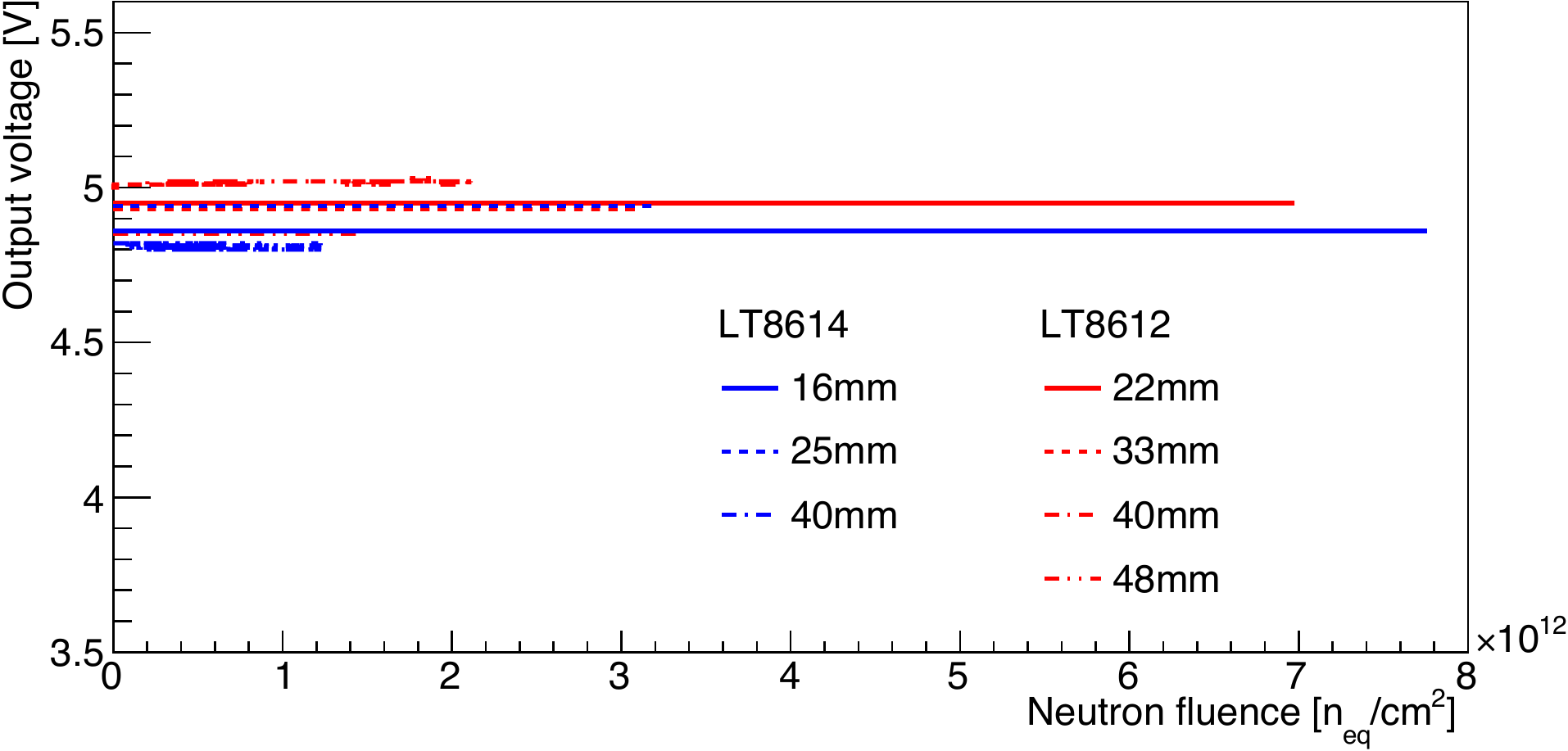}
 \end{center}
 \vspace{-5mm}
 \caption{Result of the neutron irradiation test for positive-switching regulators.}
 \label{fig:Neutron}
\end{figure}

For the positive-switching regulator selection, LT8612 and LT8614, which have enough tolerance against gamma-ray irradiation, were the focus of the neutron irradiation tests.
The flux dependence of the neutron irradiation tolerance was evaluated by placing the test boards at a different distance from the Be target, and by irradiating all of them at the same time.
Figure~\ref{fig:Neutron} shows the results.
For LT8612, the average values of neutron flux at 22\,mm and 48\,mm were $8.8\times10^7\,\neqflux\mathrm{\,sec^{-1}}$ and $1.9\times10^7\,\neqflux\mathrm{\,sec^{-1}}$, respectively.
For LT8614, the average values of neutron flux at $16\,\mathrm{mm}$ and $40\,\mathrm{mm}$ were $1.4\times10^8\,\neqflux\mathrm{\,sec^{-1}}$ and $2.3\times10^7\,\neqflux\mathrm{\,sec^{-1}}$, respectively.
All of them survived beyond the neutron dosage required and any flux dependence has not been observed.

The positive-linear regulators -- the series of LT1963 and LT3070 -- have already been used for the front-end electronics and their prototypes.
These electronics were irradiated over $1.0\times10^{12}\,\neqflux$ and permanent damage has never been observed for any of the electronics~\cite{Neutron:FPGA}.
Therefore, it was concluded that these positive-linear regulators can survive to the neutron dosage required.

\begin{figure}[htb!]
 \begin{center}
 \includegraphics[width=0.48\textwidth]{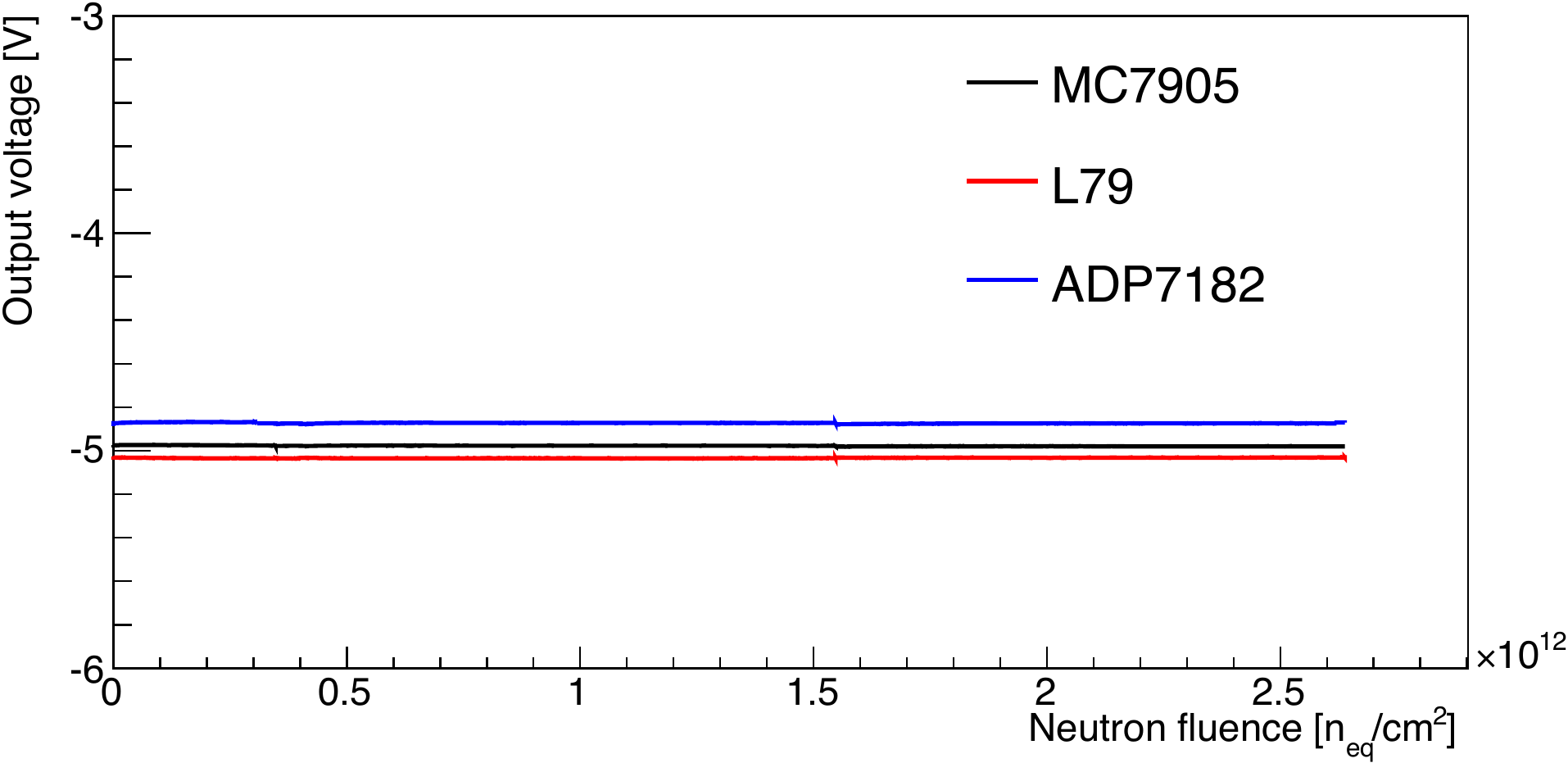}
 \end{center}
 \vspace{-5mm}
 \caption{Result of the neutron irradiation test for negative-linear regulators.}
 \label{fig:NegaLinearNeutron}
\end{figure}

Three negative-linear regulators -- L79, MC7905, and ADP7182 -- survived beyond the neutron dosage required.
Two samples for each regulator were irradiated and these results are shown in Figure~\ref{fig:NegaLinearNeutron}.
%
%

\subsection{DACs}
Ten types of DACs listed in Table~\ref{table:EvalIc} were tested for neutron tolerance, with more than $3.0 \times 10^{12} \,\neqflux$.
After the irradiation, all DACs survived, and the output voltages were consistent.

\begin{figure}[htb!]
 \begin{center}
 \includegraphics[width=0.48\textwidth]{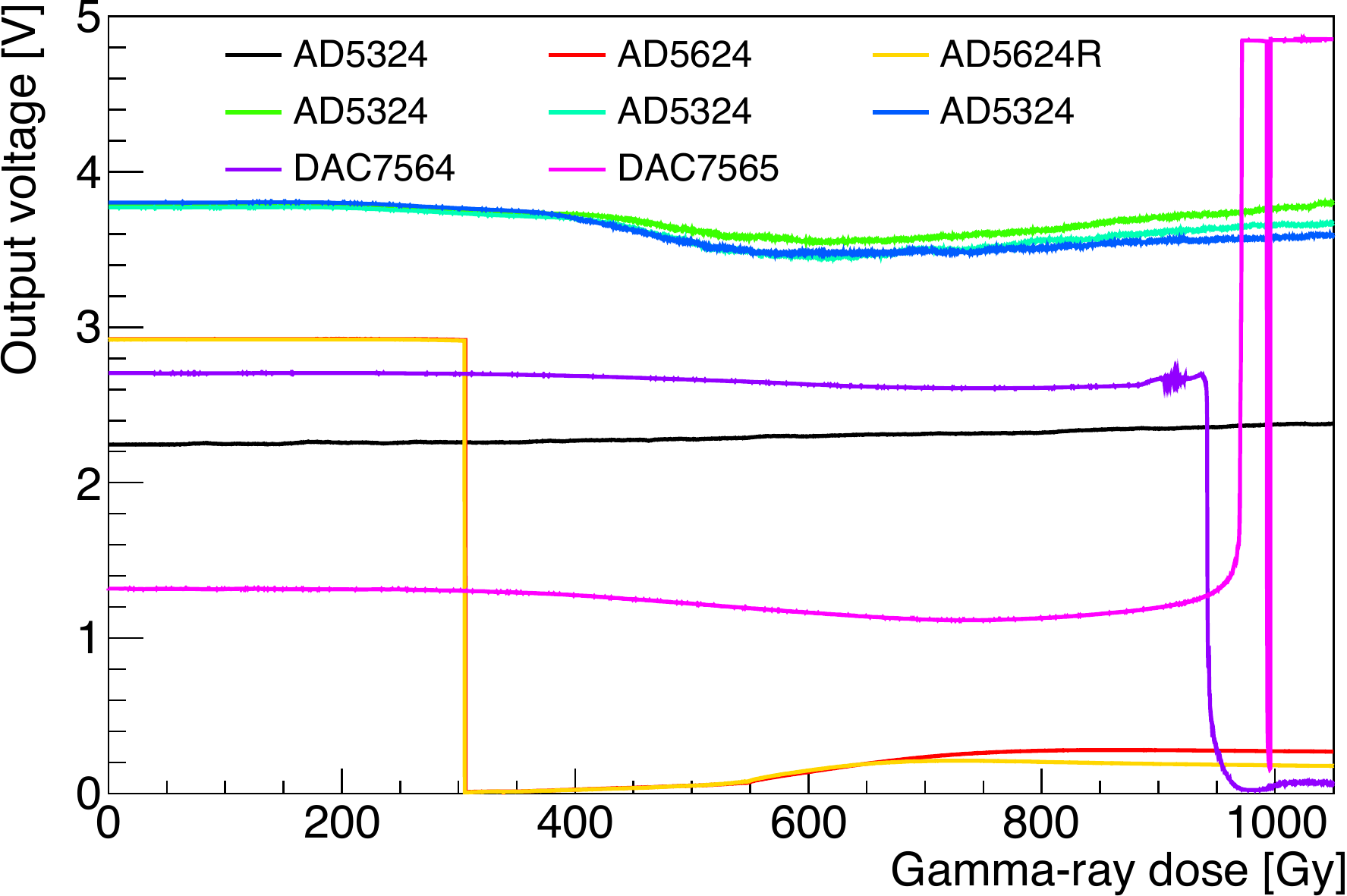}
 \end{center}
 \vspace{-5mm}
 \caption{Result of the gamma-ray irradiation test for DACs.}
 \label{fig:DAC}
\end{figure}

Out of these DACs, only six were irradiated with gamma-rays.
Figure~\ref{fig:DAC} shows the output changes in each DAC during the test.
After $0.9\,\mathrm{kGy}$ and before $1.0\,\mathrm{kGy}$ irradiation, the output of DAC7564 decreased to $0.0\,\mathrm{V}$ and the output of DAC7565 suddenly increased to $4.9\,\mathrm{V}$.
AD5624 and AD5624R stopped functioning after a power cycle after a dose of $0.3\,\mathrm{kGy}$.
Two output levels of AD5324 -- $2.2\,\mathrm{V}$ and $3.8\,\mathrm{V}$ -- were tested.
AD5324 with both output levels survived after $1.0\,\mathrm{kGy}$ irradiation while it was observed that the maximum change of its output was $0.3\,\mathrm{V}$.
%

\subsection{ADCs}
\begin{figure}[htb!]
 \vspace{-5mm}
 \centering
 \begin{minipage}{0.3\textwidth}
  \centering
  \includegraphics[width=\textwidth]{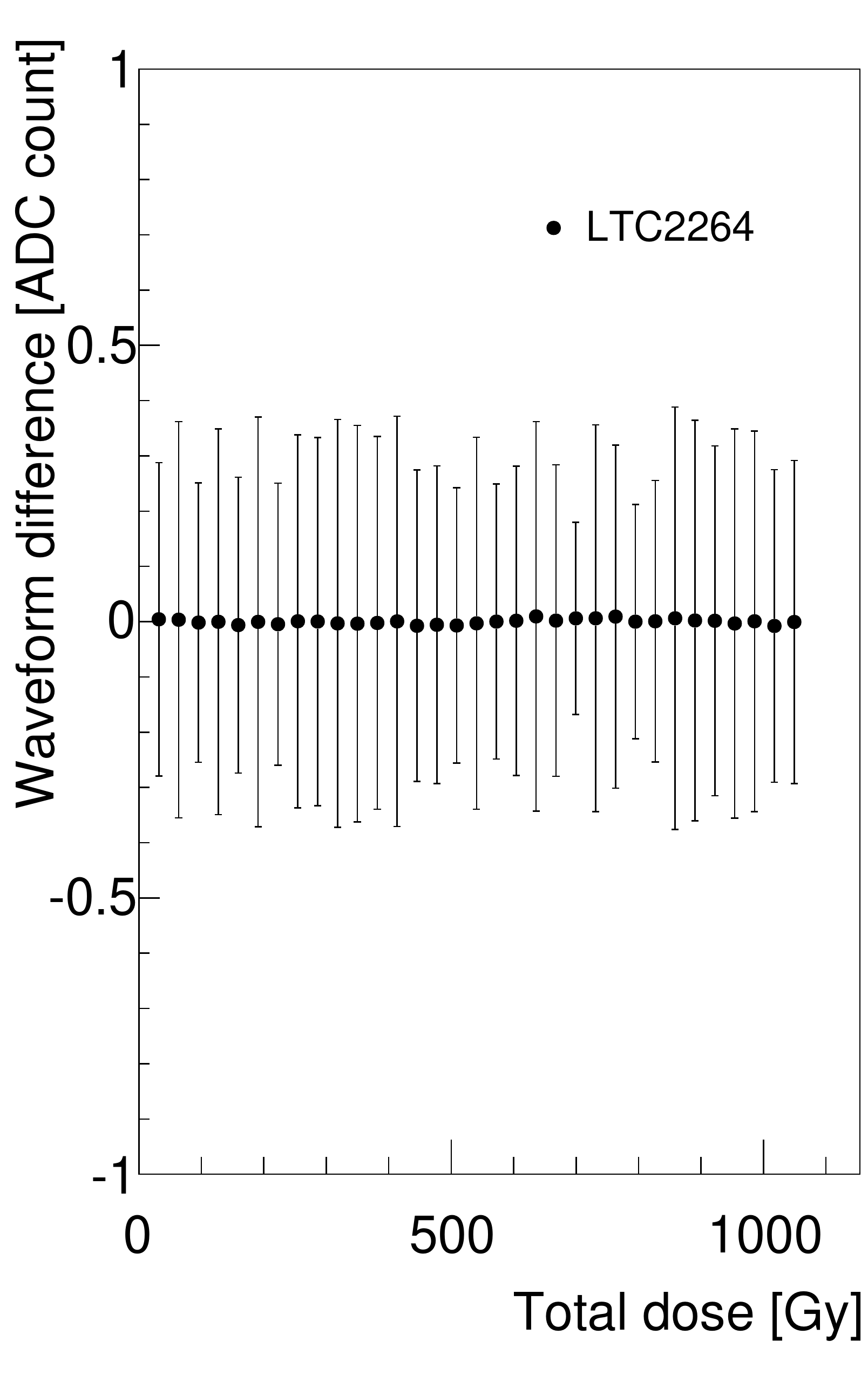}\\
  \vspace{-3mm}
  (a)
  \label{fig:LTC2264Gamma}
 \end{minipage}
 \begin{minipage}{0.3\textwidth}
  \centering
  \includegraphics[width=\textwidth]{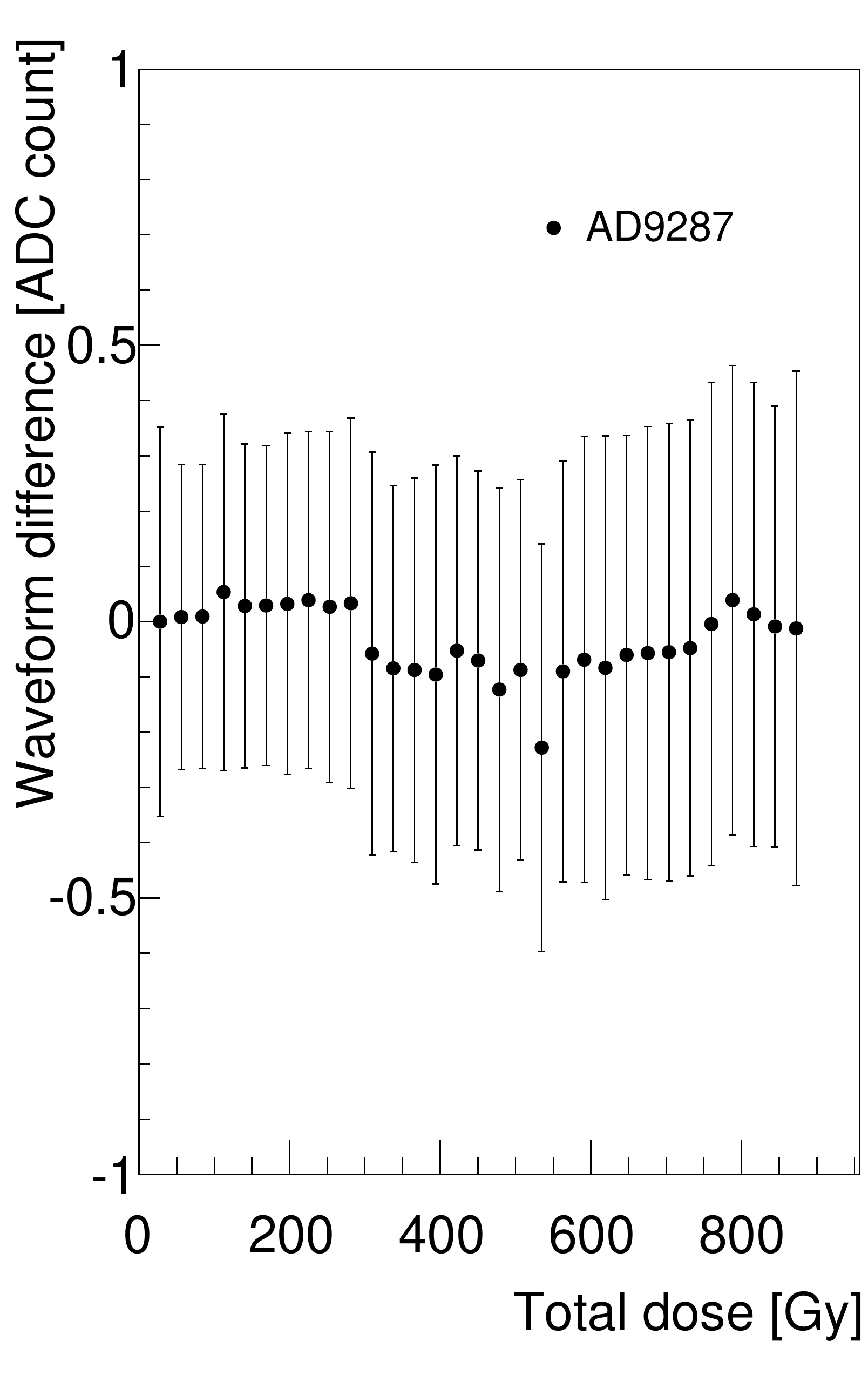}\\
  \vspace{-3mm}
  (b)
  \label{fig:AD9287Gamma}
 \end{minipage}
 \vspace{-1mm}
 \caption{Results of LTC2264 (left) and AD9287 (right) in the gamma-ray tests. These ADC outputs before and during irradiation were compared. The dots show the ratio minus 1. The error bars show root mean square of the distribution.}
 \label{fig:ADC}
\end{figure}

LTC2264 and AD9287 were irradiated with gamma-rays.
During and after irradiation, no waveform change was observed.
The results are shown in Figure~\ref{fig:ADC}.
The data points are the ratios of ADC outputs between the before and during irradiation minus 1, and the error bars are the standard deviations of the distributions.
The performance of LTC2264 was not degraded during the whole irradiation.
For AD9287 the mean values fluctuated from $0$ but did not change significantly.
Hence the performances of both the ADCs were not degraded during irradiation and after a dose of $1.0\,\mathrm{kGy}$.
AD9637 from Analog Devices was mounted on the readout electronics prototype which was irradiated with gamma-rays.
It was confirmed that it could maintain its performance after more than $1.0\,\mathrm{kGy}$ irradiation.
Power cycles for these ADCs were done after irradiation, and all of them worked fine.

Neutron tolerances of LTC2264, AD9637, and AD9287 were evaluated.
The prototype boards of the readout electronics on which these ADCs were implemented were exposed to a fluence of at least $1.0\times10^{12}\,\neqflux$~\cite{Neutron:FPGA}.
After neutron irradiation tests, the outputs of all the ADCs were unchanged.
All the ADCs exposed to the nominal neutron irradiation dose did not exhibit permanent damage.
Finally, it is concluded that the radiation tolerances of LTC2264, AD9287, and AD9637 met the requirement for the COMET Phase-I experiment.

\subsection{Other components}
After the neutron irradiation of over $1.7\times10^{12}\,\neqflux$, no deterioration was observed.
Only CD4078BM96 stopped functioning after a gamma-ray dose of $0.4\,\mathrm{kGy}$, but all the other components showed no signs of malfunction after a dose of at least $1.0\,\mathrm{kGy}$.

\subsection{Summary}
The gamma-ray and neutron irradiation tolerances for voltage regulators are summarized in Table~\ref{table:PosiReg} and Table~\ref{table:NegReg}.
Two positive-switching, six positive-linear, and three negative-linear regulators met the requirements from the COMET Phase-I radiation environment.
The result for LT1963 was consistent with the Belle II collaboration study~\cite{BelleII:Proceedings}.
Table~\ref{table:Others} shows the gamma-ray and neutron irradiation tolerances for DAC, ADC, High-Speed Differential Line Drivers, High-Speed Differential Receiver, Splitter, Positive OR gate, NOR/OR gate, and Multiplexer.
We also found candidates for the other electrical components: 1 DAC, 3 ADCs, 3 High-Speed Differential Line Drivers, 1 High-Speed Differential Receiver, 1 Splitter, 1 Positive OR gate, and 1 Multiplexer.
We plan to use Positive OR gate SN74AUP1G32 instead of NOR/OR gate.

\section{Conclusion}
Radiation tolerance of electronic components is crucial to develop reliable front-end electronics for the COMET Phase-I experiment.
Therefore, we performed a series of neutron and gamma-ray irradiation tests for the following commercial parts to investigate the radiation tolerant parts that can withstand the high radiation level expected: voltage regulators, a DAC, ADCs, High-Speed Differential Line Drivers, a High-Speed Differential Receiver, a Splitter, a Positive OR gate, a NOR/OR gate, and a Multiplexer.
The devices were exposed to gamma-rays and neutrons with the required dosage.
As a result, all of the suitable components were selected.
In addition, we will perform further irradiation tests for components where less than five samples were irradiated, to confirm our results.
We have started finalizing the designs of the electronics by using the components that passed the irradiation tests.

\section*{Acknowledgement}
The authors are grateful to Prof. Y. Furuyama, Dr. A. Taniike, Mr. T. Yokose and Mr. H. Kageyama (Kobe University) for the operation of the tandem electrostatic accelerator; Prof. K. Sagara, Prof. T. Kin, Prof. S. Sakaguchi and Dr. S. Araki (Kyushu University) for the operation of the tandem electrostatic accelerator; Dr. A. Idesaki (QST) for gamma-ray exposure in National Institute for Quantum and Radiological Science and Technology; Mr. I. Yoda (Tokyo Institute of Technology) for gamma-ray exposure in Radioisotope Research Center; Dr. S. Tojo and Mr. Y. Okada (Osaka University) for gamma-ray exposure in ISIR.
This work was supported by JSPS KAKENHI Grant Numbers JP17H04841, JP25000004, JP18H03704 and JP18H05231; Institute for Basic Science (IBS) of Republic of Korea under Project No. IBS-R017- D1-2018-a00; National Natural Science Foundation of China (NSFC) under Contracts No. 11335009.


\begin{table*}[htb!]
\centering
\caption{Summary of gamma-ray and neutron irradiation tolerances for positive-voltage regulators. Loads are the resistances of resistors connected to the regulator outputs. The values of input voltage were measured on the input ports of the test boards. In the column of the neutron tolerance blanks mean no measurement in the tests. The neutron tolerances of the series of LT1963, LT3070 and LMZ10503 were calculated from the results of the neutron irradiation tests for the front-end electronics, where these regulators were mounted~\cite{Neutron:FPGA}.}
\label{table:PosiReg}
\begin{minipage}{\textwidth}
 \begin{center}
 \small
 \begin{adjustbox}{width=\textwidth,totalheight=\textheight,keepaspectratio}
 \renewcommand{\thefootnote}{\arabic{footnote}}
 \begin{tabular}{lllcccc}
 \hline \hline
  & & & & & \multicolumn{2}{c}{\shortstack{Tolerance}} \\
  \raisebox{0.5em}[0pt][0pt]{Type} & \raisebox{0.5em}[0pt][0pt]{Manufacturer} & \raisebox{0.5em}[0pt][0pt]{Name} & \raisebox{0.5em}[0pt][0pt]{Load $\left(\mathrm{\Omega}\right)$} & \raisebox{0.5em}[0pt][0pt]{Input Voltage $\left(\mathrm{V}\right)$} & Gamma-ray $\left(\mathrm{kGy}\right)$ & Neutron $\left(\neqflux\right)$ \\ \hline
  Linear    & Linear Technology & LT1963     & $6.2$ & $1.9$ & $>2.4$ & $>7.2\times 10^{12}$ \\ 
 			& 					& LT1963-3.3 & $13$  & $3.7$ & $>3.4$ & $>7.4\times 10^{12}$ \\
 			&					& LT1963-2.5 & $10$  & $2.9$ & $>3.4$ & $>7.5\times 10^{12}$ \\
 			&					& LT1963-1.8 & $7.5$ & $2.2$ & $>3.4$ & $>6.3\times 10^{12}$ \\
            &					& LT3070	 & $1.6$ & $1.3$ & $>2.0$ & $>8.0\times 10^{12}$ \\
            &					& LT1764A	 & $3.0$ & $1.9$ & $<0.1$ & \\
            & 					& LTC3026	 & $10$  & $1.5$ & $1.4$  & $>3.1\times 10^{12}$ \\ \cmidrule(r){2-7} 
            & Maxim Integrated	& MAX8556	 & $1.6$, $0.53$ \footnote[1]{5 samples were tested with each load.} & $1.4$ & $0.8$  & \\ \cmidrule(r){2-7}
            & Texas Instruments	& TPS7A7200	 & $10$  & $3.3$ & $1.0$  & \\
            &					& TPS75801	 & $10$. $33$, $120$ \footnote[2]{2 samples were tested with each load.} & $3.3$ & $0.6$  & \\
            &					& TPS74401	 & $10$  & $3.3$ & $1.4$  & \\
            & Analog Devices    & ADP1755    & $10$  & $3.3$ & $0.8$ \footnote[3]{4 samples were tested with dose rate of $2.3\times10^2\,\mathrm{Gy\,h^{-1}}$} & \\ \hline
  Switching	& Linear Technology	& LT8612     & $39$  & $6.0$ & $>2.2$ & $>7.0\times 10^{12}$ \\ 
            &					& LT8614     & $39$  & $6.0$ & $>2.2$ & $>7.8\times 10^{12}$ \\
            &					& LTM4620	 & $1.0$, $6.2$ \footnote[4]{$1.0\,\mathrm{\Omega}$ load and $6.2\,\mathrm{\Omega}$ load were used for outputs of $1.0\,\mathrm{V}$ and $2.5\,\mathrm{V}$, respectively.}  & $5.0$ & $0.3$  & \\
            &					& LTM4644	 & $1.6$, $2.0$, $2.7$, $3.0$ \footnote[5]{4 loads were adopted for 4 outputs from LTM4644: $1.6\,\mathrm{\Omega}$ for $1.0\,\mathrm{V}$ output, $2.0\,\mathrm{\Omega}$ for $1.2\,\mathrm{V}$ output, $2.7\,\mathrm{\Omega}$ for $1.5\,\mathrm{V}$ output, and $3.0\,\mathrm{\Omega}$ for $1.8\,\mathrm{V}$ output.} & $5.0$ & $1.1$  & \\
            &					& LTM8033	 & $38$  & $6.0$ & $0.2$  & \\ \cmidrule(r){2-7} 
            & Texas Instruments & LMZ10503	 & $2.7$ & $5.0$ & $0.8$  & $>2.8\times 10^{12}$ \\ \hline
 \end{tabular}
 \end{adjustbox}
 \end{center}
\end{minipage}
\end{table*}

\begin{table*}[htb!]
\centering
\caption{Summary of gamma-ray and neutron irradiation tolerances for negative-voltage regulators. Loads are the resistances of resistors connected to the regulator outputs. Input voltages are the setup value of power supply modules. In the column of the neutron tolerance blanks mean no measurement in the tests.}
\label{table:NegReg}
\begin{minipage}{\textwidth}
 \begin{center}
 \small
 \begin{adjustbox}{width=\textwidth,totalheight=\textheight,keepaspectratio}
 \begin{tabular}{lllcccc} \hline \hline
  & & & & & \multicolumn{2}{c}{\shortstack{Tolerance}} \\
  \raisebox{0.5em}[0pt][0pt]{Type} & \raisebox{0.5em}[0pt][0pt]{Manufacturer} & \raisebox{0.5em}[0pt][0pt]{Name} & \raisebox{0.5em}[0pt][0pt]{Load $\left(\mathrm{\Omega}\right)$} & \raisebox{0.5em}[0pt][0pt]{Input Voltage $\left(\mathrm{V}\right)$} & Gamma-ray $\left(\mathrm{kGy}\right)$ & Neutron $\left(\neqflux\right)$ \\ \hline
  Linear	& ST Microelectronics	& L79		 & $47$ & $-7.0$ & $>2.0$ & $>2.6\times 10^{12}$ \\ \cmidrule(r){2-7} 
 			& ON Semiconductor		& MC7905	 & $47$ & $-7.0$ & $>2.0$ & $>2.6\times 10^{12}$ \\ \cmidrule(r){2-7} 
 			& New JRC				& NJM2828	 & $47$ & $-7.0$ & $>2.0$ & \\ \cmidrule(r){2-7} 
 			& Linear Technology		& ADP7182	 & $47$ & $-7.0$ & $>2.0$ & $>2.6\times 10^{12}$ \\
            &						& LT1964	 & $47$ & $-7.0$ & $<0.1$ & \\
            &						& LT3015	 & $47$ & $-7.0$ & $0.6$  & \\	
            &						& LT3032	 & $47$ & $-7.0$ & $<0.1$ & \\
            &						& LT3090	 & $47$ & $-7.0$ & $<0.1$ & \\
            &						& LT3091	 & $47$ & $-7.0$ & $<0.1$ & \\ \cmidrule(r){2-7} 
            & Texas Instruments		& LM337		 & $47$ & $-7.0$ & $0.4$  & $>2.6\times 10^{12}$ \\ \cmidrule(r){2-7}
            & Microchip Technology	& MIC5271	 & $47$ & $-7.0$ & $<0.1$ & $>2.6\times 10^{12}$ \\
            &						& TC59		 & $47$ & $-7.0$ & $0.2$  & $>2.6\times 10^{12}$ \\ \hline
 \end{tabular}
 \end{adjustbox}
 \end{center}
\end{minipage}
\end{table*}

\begin{table*}[htb!]
\centering
\caption{Summary of gamma-ray and neutron irradiation tolerances for DAC, ADC, High-Speed Differential Line Driver, High-Speed Differential Receiver, Splitter, Positive OR gate, NOR/OR gate, and Multiplexer. In the column of the gamma-ray irradiation tolerance blanks mean no measurement in the tests.}
\label{table:Others}
\begin{tabular}{cc}
 \begin{minipage}{0.49\textwidth}
  \begin{center}
  \small
  \begin{adjustbox}{width=\textwidth,totalheight=\textheight,keepaspectratio}
  \begin{tabular}{llcc} \hline \hline
  & & \multicolumn{2}{c}{\shortstack{Tolerance}} \\
  \raisebox{0.5em}[0pt][0pt]{Manufacturer} & \raisebox{0.5em}[0pt][0pt]{Name} & Gamma-ray $\left(\mathrm{kGy}\right)$ & Neutron $\left(\neqflux\right)$ \\ \hline
  \multicolumn{4}{c}{\shortstack{\bf{DAC} \rule[-3mm]{0mm}{7mm} }} \\
  Linear Technology     & LTC2624   & $0.9$     & $>3.0\times 10^{12}$ \\
                        & LTC2634   &           & $>3.0\times 10^{12}$ \\
                        & LTC2654   &           & $>3.0\times 10^{12}$ \\
  Texas Instruments     & DAC7564   & $0.9$     & $>3.0\times 10^{12}$ \\
                        & DAC7565   & $0.9$     & $>3.0\times 10^{12}$ \\
  Analog Devices        & AD5624R   & $0.3$     & $>3.0\times 10^{12}$ \\
                        & AD5684R   &           & $>3.0\times 10^{12}$ \\
                        & AD5324    & $>1.0$    & $>3.0\times 10^{12}$ \\
                        & AD5624    & $0.3$     & $>3.0\times 10^{12}$ \\
                        & AD5684    &           & $>3.0\times 10^{12}$ \\ \hline
  \multicolumn{4}{c}{\shortstack{\bf{ADC} \rule[-3mm]{0mm}{7mm} }} \\
  Linear Technology     & LTC2264   & $>1.0$    & $>1.0\times 10^{12}$ \\
  Analog Devices        & AD9287    & $>1.0$    & $>1.0\times 10^{12}$ \\
                        & AD9637    & $>1.0$    & $>1.0\times 10^{12}$ \\ \hline
  \end{tabular}
  \end{adjustbox}
  \end{center}
 \end{minipage}
 \begin{minipage}{0.49\textwidth}
  \begin{center}
  \small
  \begin{adjustbox}{width=\textwidth,totalheight=\textheight,keepaspectratio}
  \begin{tabular}{llcc} \hline \hline
  & & \multicolumn{2}{c}{\shortstack{Tolerance}} \\
  \raisebox{0.5em}[0pt][0pt]{Manufacturer} & \raisebox{0.5em}[0pt][0pt]{Name} & Gamma-ray $\left(\mathrm{kGy}\right)$ & Neutron $\left(\neqflux\right)$ \\ \hline
  \multicolumn{4}{c}{\shortstack{\bf{High-Speed Differential Line Driver} \rule[-2.5mm]{0mm}{6mm} }} \\
  Texas Instruments     & SN65LVDS391   & $>1.0$    & $>1.7\times 10^{12}$ \\
                        & SN65LVDS116   & $>1.0$    & $>1.7\times 10^{12}$ \\
                        & SN65LVDS386   & $>1.0$    & $>1.7\times 10^{12}$ \\ \hline
  \multicolumn{4}{c}{\shortstack{\bf{High-Speed Differential Receiver} \rule[-2.5mm]{0mm}{6mm} }} \\
  Texas Instruments     & SN65LVDT348   & $>1.0$    & $>1.7\times 10^{12}$ \\ \hline
  \multicolumn{4}{c}{\shortstack{\bf{Splitter} \rule[-2.5mm]{0mm}{6mm} }} \\
  Maxim Integrated      & MAX9175       & $>1.0$    & $>1.7\times 10^{12}$ \\ \hline
  \multicolumn{4}{c}{\shortstack{\bf{Positive OR gate} \rule[-2.5mm]{0mm}{6mm} }} \\
  Texas Instruments     & SN74AUP1G32   & $>1.0$    & $>1.7\times 10^{12}$ \\ \hline
  \multicolumn{4}{c}{\shortstack{\bf{NOR/OR gate} \rule[-2.5mm]{0mm}{6mm} }} \\
  Texas Instruments     & CD4078BM96    & $0.4$     & $>1.7\times 10^{12}$ \\ \hline
  \multicolumn{4}{c}{\shortstack{\bf{Multiplexer} \rule[-2.5mm]{0mm}{6mm} }} \\
  Analog Devices        & ADG1606       & $>1.0$    & $>1.7\times 10^{12}$ \\ \hline
  \end{tabular}
  \end{adjustbox}
  \end{center}
  \normalsize
 \end{minipage}
\end{tabular}
\end{table*}

\end{document}